\RequirePackage{fix-cm}

\documentclass[smallextended]{svjour3}       % onecolumn (second format)
\usepackage{pgfplots}
\usepackage{pgfplotstable}
\usepackage{tikz}
\usepackage{booktabs}
\usepackage{adjustbox}
\usepackage{xcolor}
\usetikzlibrary{arrows,shadows,backgrounds,calc,decorations.pathreplacing,decorations.pathmorphing}
   \definecolor{webgreen}{rgb}{0,.5,0}
    \definecolor{webblue}{rgb}{0,0,.8}
    \definecolor{webred}{rgb}{0.8, 0, 0}   
   \definecolor{webbrown}{rgb}{.6,0,0}
    \definecolor{webyellow}{rgb}{0.98,0.92,0.73}
    \definecolor{webgray}{rgb}{.753,.753,.753}
\newcommand{\gls}[1]{{\bfseries#1}}
\def\scalefact{.8}
\tikzstyle{QT} = [top color=white, bottom color=blue!30,
minimum height={0.8},rotate=90,scale=\scalefact,xshift=-.5cm,yshift=-.5cm,
draw=blue!50!black!100, drop shadow] %, font=\small]

\tikzstyle{QTC} = [top color=green!80, bottom color=white,
minimum height={0.8},scale=\scalefact,xshift=-.5cm,yshift=-.8cm,
draw=blue!50!black!100, drop shadow] %, font=\small]

\pgfplotscreateplotcyclelist{my color list}{%
solid, color=webblue, every mark/.append style={solid, fill=webblue}, mark=*\\%
densely dashdotted, color=webgreen, every mark/.append style={solid, fill=webred},mark=diamond*\\%
densely dotted, color=webbrown, every mark/.append style={solid, fill=webgreen}, mark=triangle*\\%
loosely dashed, color=webred, every mark/.append style={solid, fill=webbrown},mark=*\\%
dotted, color=webblue, every mark/.append style={solid, fill=webyellow}, mark=square*\\%
densely dotted, color=webgreen, every mark/.append style={solid, fill=gray}, mark=otimes*\\%
dashed, color=webbrown, every mark/.append style={solid, fill=gray},mark=diamond*\\%
densely dashed, every mark/.append style={solid, fill=gray},mark=square*\\%
dashdotted, every mark/.append style={solid, fill=gray},mark=otimes*\\%
dashdotdotted, every mark/.append style={solid},mark=star\\%
}
\tikzstyle{QTC} = [top color=green!80, bottom color=white,
minimum height={0.8},scale=\scalefact,xshift=-.5cm,yshift=-.8cm,
draw=blue!50!black!100, drop shadow] %, font=\small]
\newcommand\QTfigure[6]   
{
	\draw
	($ (0 cm,0.5cm-.5*#2 cm) + (0 cm,-#5*.1 cm) +(#3cm,-#4cm)$) node[QT,minimum width=#2cm,draw]
	(Q#1) {$#1$}
	;
} 
\newcommand\QTCfigure[6]   
{
	\draw
	($ (2.5 cm,.6cm-.5*#2 cm) + (0.5 cm,-#5*.1 cm) +(#3cm,-#4cm)$) node[QTC,minimum width=5cm,minimum height=#2cm,draw]
	(Q#1) {$#1$}
	;
} 
\newcommand\QTAfigure[6]   
{
	\draw
	($ (3 cm
	,%.6cm
	.6cm-.5*#2 cm) + (0 cm,-#5*.1 cm) +(#3cm,-#4cm)$) node[QTC,minimum width=3cm,minimum height=#2cm,draw]
	(Q#1) {${#1}$}
	;
} 

\newcommand\QTRfigure[6]   
{
	\draw
	($ (0 cm,0.5cm-.5*#2 cm) + (0 cm,-#5*.1 cm) +(#3cm,-#4cm)$) node[QT,color=red,minimum width=#2cm,draw]
	(Q#1) {$#1$}
	;
} 

\smartqed  % flush right qed marks, e.g. at end of proof
\usepackage{graphicx}
\begin{document}

\title{Statistical considerations on limitations of supercomputers\thanks{Project no. 125547 has been implemented with the support provided from the National Research, Development and Innovation Fund of Hungary, financed under the K funding scheme.}
}
%\subtitle{Based on the %publicly available 
%TOP500 database}

\titlerunning{Statistical considerations of supercomputers}        % if too long for running head

\author{J\'anos V\'egh
}


\institute{J. V\'egh \at
              University of Miskolc \\
              Tel.: +36-46-565-111/1753\\
              \email{J.Vegh@uni-miskolc.hu}           %  \\
}

\date{Received: date / Accepted: date}

\maketitle

\begin{abstract}
Supercomputer building is a many sceene, many authors game,
comprising a lot of different technologies, manufacturers and ideas. 
Checking data available in the public database in a  systematic way,
some general tendencies and limitations can be concluded, both for the past and the future.
The feasibility of building exa-scale computers as well as  their limitations and utilization are also discussed.
The statistical considerations provide a strong support for the conclusions.

\end{abstract}

\section{Introduction}
\label{intro}
For now, supercomputing has a quarter of century history and a well-documented 
and verified database~\cite{Top500:2016} on their architectural and performance data.
The huge variety of solutions and ideas does not enlighten drawing conclusions and especially making forecasts for the future of supercomputing.

In section~\ref{sec:Amdahl} Amdahl's law is reconsidered, interpreting it for the modern computing architectures,
with keeping an eye on measurability. 
Choosing the right merit~\cite{Vegh:2017:AlphaEff} of their characteristics and 
utilizing a large number of reliable measured data~\cite{Top500:2016},  clear conclusions are drawn in section~\ref{sec:checks}.
After validating the method, some predictions are made through extrapolating the tendencies for the near future in section~\ref{sec:future}.

\section{Supercomputers and Amdahl's law}\label{sec:Amdahl}

Amdahl's law~\cite{AmdahlSingleProcessor67} on the joint performance of parallelly working systems "is one of the few, fundamental laws of computing"~\cite{AmdalsLaw-Paul2007} which seems to be nearly forgotten in the field of supercomputing.
As tought in introductory courses on parallel processing, some fraction of the computing job cannot be parallelized (i.e. cannot be distributed among the parallelly working units),
and this fraction limits the achievable resulting computing performance.

Although Amdahl only wanted to draw the attention to that the so called \textit{Single-Processor Approach} introduces some serious limitations
on computing performance (especially when large number of processors is utilized in the system of parallelly working processors),
his successors  formulated his idea (commonly known as \textit{Amdahl's law}) differently.
A common misconception is to assume that Amdahl's law
is valid for software only and that parallelizable fraction $\alpha$ contains
something like ratio of numbers of the corresponding instructions to the respective total number.

Amdahl's law is much more general, and is actually used on many different fields~\cite{UsesAbusesAmdahl:2001}. If Amdahl's law is interpreted correctly: for \textit{the time needed for some activity} rather than for some \textit{fraction of the code},
it should describe also performance and limits of operation of supercomputers.
Even, supercomputers are an excellent playground to check validity of Amdahl's law in the case of extremely large number of processors.
% To do so, the non parallelizable fraction should be interpreted properly.

\subsection{Terms in Amdahl's law}
First the notations used in~\cite{Vegh:2017:AlphaEff} are introduced
and a summary of the ideas explained and illustrated in details there
is given.
If $\alpha$ stands for \textit{the time fraction} of activity that can be outsourced to several 
parallelly working processing units, all the rests, $(1-\alpha)$ fraction, independently of their origin, fall into the category of non-parallelizable activity and (as discussed by Amdahl) \textit{appear as if they were sequential-only} activity. 
If the parallelizable fraction is distributed among $k$ processing units,
the speedup $S$ which can be achieved is

\begin{equation}
S^{-1}=(1-\alpha) +\alpha/k \label{eq:AmdahlBase}
\end{equation}

The speedup multiplied with the $P$ absolute performance of one processor, the (apparent) resulting performance\footnote{The factor $\frac{1}{(1-\alpha)}$ can be considered as a kind of \textit{performance gain} or performance amplification factor} is given as

\begin{equation}
P_{Max}=P\frac{1}{(1-\alpha)}\label{eq:PerformanceMax}
\end{equation}

This is a \textit{theoretical upper limit} for the performance (also of a supercomputer) which can only be achieved in idealistic case,
as discussed in~\cite{Vegh:2017:AlphaEff}. This usually cannot be computed in advance,
because $\alpha$ is not known in advance.
However, on a "black box" supercomputer one can measure $R_{Max}$
and it is also known that $R_{Peak}=kP$. Since 

\begin{equation}
S=\frac{(1-\alpha)+\alpha}{(1-\alpha)+\alpha/k} =\frac{k}{k(1-\alpha)+\alpha}
\end{equation}

\noindent and the efficiency

\begin{equation}
E = \frac{S}{k}=\frac{1}{k(1-\alpha)+\alpha}=\frac{R_{Max}}{R_{Peak}}\label{eq:soverk}
 \end{equation}

\noindent the measured payload performance provides information also on the "effective parallelism".
That is, only a fraction of nominal performance can be utilized
as payload performance, the rest remains a kind of "\textit{dark performance}".
\noindent One can easily express the "effective parallelization" $\alpha_{eff}$
from the measured efficiency as

\begin{equation}
\alpha_{eff} = \frac{k}{k-1}\frac{S-1}{S} \label{equ:alphaeff}
\end{equation}
or equivalently
\begin{equation}
 \alpha_{eff} = \frac{E k -1}{E (k-1)}\label{eq:alphafromE}
 \end{equation}
Using measured performance values published for supercomputers~\cite{Top500:2016},
$\alpha_{eff}$ values for the supercomputer configurations can be calculated, see Fig~\ref{SupercomputerTimeline}.
Notice that for a given configuration $\alpha_{eff}$  depends on $k$.

\subsection{A simple model for supercomputing}

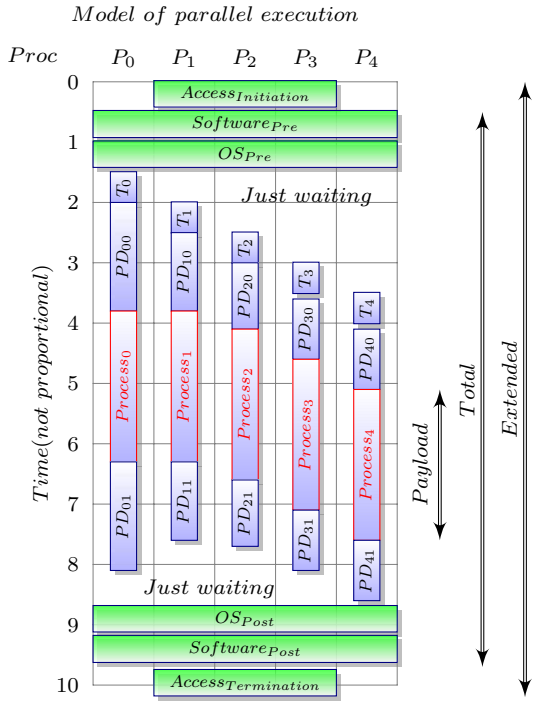
\begin{figure} %[!b]
	\begin{tikzpicture}[scale=.8,cap=round]
	\node[right,above] at (11cm,.2cm) {$Proc$};
	\node[rotate=90,above=.5] at (11.5cm,-5cm) {$Time (not\  proportional)$};
	\foreach \y/\ytext in {0,...,10}
	\draw[yshift=-\y cm,xshift=-0.1cm+12cm]  node[left]
	{${\ytext}$};
	\draw[style=help lines,step=1] (12,-10) grid (17,0);
	\node[right,above] at (14cm,.8cm) {$Model\ of\ parallel\ execution$};
	\foreach \x/\xtext in {0,...,4}
	\draw[xshift=\x cm+12.5 cm,yshift=0.1cm]  node[above=.1]
	{$P_{\xtext}$};
	\QTAfigure{Access_{Initiation}}{.0}{12}{0}{0}{0}  % 
	\QTCfigure{Software_{Pre}}{.0}{12}{0}{5}{0}  % 
	\QTCfigure{OS_{Pre}}{.0}{12}{0}{10}{0}  % 
	\QTfigure{T_0}{0.5}{12}{0}{15}{0}  % 
	\QTfigure{PD_{00}}{1.8}{12}{0}{20}{0}  % 
	\QTRfigure{Process_0}{2.5}{12}{0}{38}{0}  % 
	\QTfigure{PD_{01}}{1.8}{12}{0}{63}{0}  % 
	\QTfigure{T_1}{0.5}{13}{0}{20}{0}  % 
	\QTfigure{PD_{10}}{1.3}{13}{0}{25}{0}  % 
	\QTRfigure{Process_1}{2.5}{13}{0}{38}{0}  % 
	\QTfigure{PD_{11}}{1.3}{13}{0}{63}{0}  % 
	\QTfigure{T_2}{0.5}{14}{0}{25}{0}  % 
	\QTfigure{PD_{20}}{1.1}{14}{0}{30}{0}  % 
	\QTRfigure{Process_2}{2.5}{14}{0}{41}{0}  % 
	\QTfigure{PD_{21}}{1.1}{14}{0}{66}{0}  % 
	\QTfigure{T_3}{0.5}{15}{0}{30}{0}  % 
	\QTfigure{PD_{30}}{1}{15}{0}{36}{0}  % 
	\QTRfigure{Process_3}{2.5}{15}{0}{46}{0}  % 
	\QTfigure{PD_{31}}{1}{15}{0}{71}{0}  % 
	\QTfigure{T_4}{0.5}{16}{0}{35}{0}  % 
	\QTfigure{PD_{40}}{1}{16}{0}{41}{0}  % 
	\QTRfigure{Process_4}{2.5}{16}{0}{51}{0}  % 
	\QTfigure{PD_{41}}{1}{16}{0}{76}{0}  % 
	\node[right,above] at (15.5cm,-2.2cm) {$Just\ waiting$};
	\node[right,above] at (13.9cm,-8.7cm) {$Just\ waiting$};

	\QTCfigure{OS_{Post}}{.0}{12}{0}{87}{0}  % 
	\QTCfigure{Software_{Post}}{.0}{12}{0}{92}{0}  % 
	\QTAfigure{Access_{Termination}}{.0}{12}{0}{97.6}{0}  % 
	
	%
	%\draw[latex'-latex',double] (B) -- node[label=90:B-C,label=270:C-B] {} (C);
	\coordinate (TP1) at (17.7 cm, -5.1);
	\coordinate (TP2) at (17.7 cm, -7.6);
	\draw[latex'-latex',double] (TP1) -- node[above, rotate=90] {$Payload$} (TP2);
	
	\coordinate (TT1) at (18.4 cm, -0.5);
	\coordinate (TT2) at (18.4 cm, -9.7);
	\draw[latex'-latex',double] (TT1) -- node[above,
	rotate=90] {$Total$} (TT2);
	
	\coordinate (TE1) at (19.1 cm, 0);
	\coordinate (TE2) at (19.1 cm, -10.2);
	\draw[latex'-latex',double] (TE1) -- node[above,
	rotate=90] {$Extended$} (TE2);
	\end{tikzpicture}
\caption{The extended Amdahl's model (somewhat idealistic)}
\label{fig:Ourmodel}{}{}
\end{figure}

To understand the
meaning of the values derived in this way, a simple model  shown in Fig.~\ref{fig:Ourmodel} should be derived. 
Although the model is empirical rather than technical,
with slightly extending it and giving technical meaning to its terms,
it can easily be converted to technical model.
Also note that here no communication is assumed between the parallelly working units, but the model can be trivially extended to the case when the parallelly working processors communicate (explicitly or implicitly, like sharing some resource). The model assumes that several components contribute to the 
total execution time,  as simple sum of either some components or the largest of some components.

 The access time is usually small: whether the time
is measured on the parallelized system or outside of it, one must compensate for its contribution (in the case of supercomputers, it is usually negligible).
The contribution of the executed program $\alpha_{eff}^{SW}$
depends heavily on the nature of the program.
The contributions due to \gls{OS} and \gls{HW} are tightly connected,
so it is not easy to separate them without making dedicated measurements;
at this level their joint contribution will be handled as  $\alpha_{eff}^{HW+OS}$.
Within that contributions there are some parts which may become critical,
like the looping delay $T_x$ due to utilizing extremely large number of processors or the propagation delay $PD_{xx}$ due to having large physical size of the supercomputer;
they will be mentioned separately, and in the technical model they shall be handled specifically.
The time scale shown in the figure serves only for illustration, the actual contributions will strongly vary with the actual conditions.

From the figure the meaning of $\alpha_{eff}$ can be easily identified as $Payload/Total$. Also, the reasons of "dark performance" can be identified: the ready-to-fire processing units are simply idle. The common mistake of handling the access time improperly can falsify the conclusions, although in the case of long measurement times this effect can be neglected.

\section{Performance and architecture checks}\label{sec:checks}

The available, rigorously validated database~\cite{Top500:2016} enables to draw reliable conclusions, although  the variety of sources of components, different technologies and ideas as well as the interplay of different factors cause a considerable scatter
and requires extremely careful analysis.

\subsection{Supercomputer timeline}
\begin{figure*}
\begin{tikzpicture}[scale=.95]
\begin{axis}
[
	title={Supercomputers, Top 500 1st-3rd},
	width=\textwidth,
	cycle list name={my color list},
		legend style={
			cells={anchor=east},
			legend pos={north east},
		},
		xmin=1993, xmax=2017,% x scale
		ymin=1e-8, ymax=1e-2, % y scale
		xlabel=Year,
		/pgf/number format/1000 sep={},
		ylabel=$(1-\alpha)$,
		ymode=log,
		log basis x=2,
		]
\addplot table [x=a, y=b, col sep=comma] {Top500-0.csv};
		\addlegendentry{$1st $}
\addplot table [x=a, y=c, col sep=comma] {Top500-0.csv};
		\addlegendentry{$2nd $}
\addplot table [x=a, y=d, col sep=comma] {Top500-0.csv};
		\addlegendentry{$3rd $}
\addplot table [x=a, y=e, col sep=comma] {Top500-0.csv};
		\addlegendentry{$Best\ \alpha$}
		\addplot[ very thick, color=webbrown] plot coordinates {
			(1993, 1e-3)  
			(2017,1e-7) 
		};
		\addlegendentry{Trend of $(1-\alpha)$}
		\addplot[only marks, color=red, mark=star,  mark size=3, very thick] plot coordinates {
			(2016,33e-9) 
		};
		\addlegendentry{Sunway TaihuLight}
\end{axis}
\end{tikzpicture}
\caption{The trend of the development of ($1-\alpha$) in the past 25 years, based on the
first three (by $R_{max}$) and the first (by ($1-\alpha$)) in the year in question.}
\label{SupercomputerTimeline}
\end{figure*}
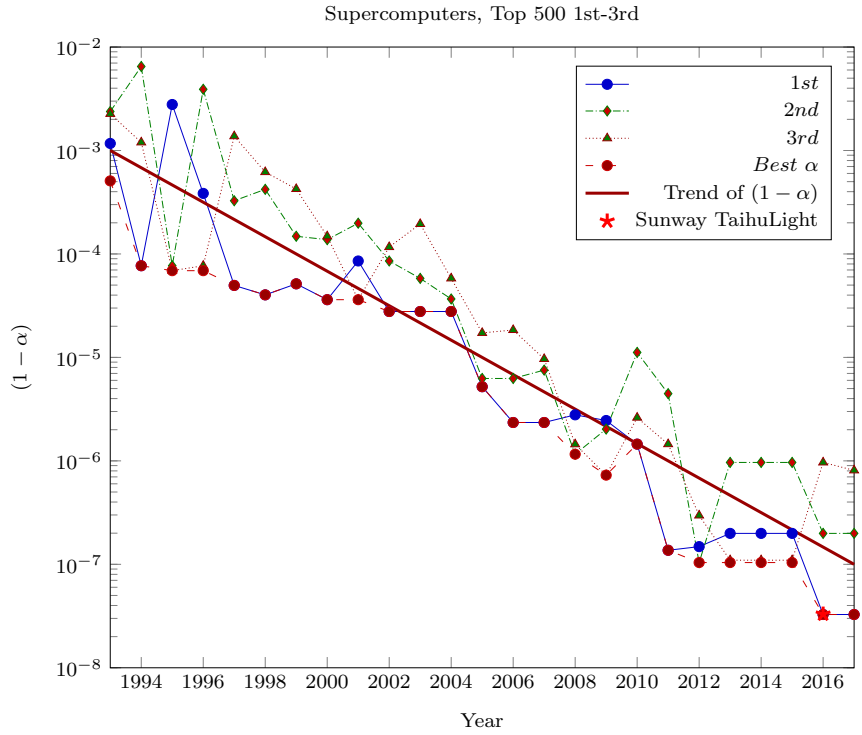

As a quick test,  Equ.~(\ref{eq:alphafromE}) can be applied to data from~\cite{Top500:2016}, see Fig.~\ref{SupercomputerTimeline}. 
As shown, supercomputer history is about the development of effective parallelism,
and Amdahl's law formulated by Equ.~(\ref{eq:alphafromE}) 
is actually what Moore's law is for the size of electronic components.
(The effect of Moore's law is eliminated when calculating $\frac{R_{Max}}{R_{Peak}}$.) To understand the behavior of the trend line,
just recall Equ.~(\ref{eq:soverk}): to increase the absolute performance,
more processors shall be included, and to provide reasonable efficiency,
the value of $(1-\alpha)$ must be properly reduced.
Just notice that the excellent performance of $Taihulight$
shall be attributed to its special processor, deploying "Cooperative computing"~\cite{CooperativeComputing2015}.

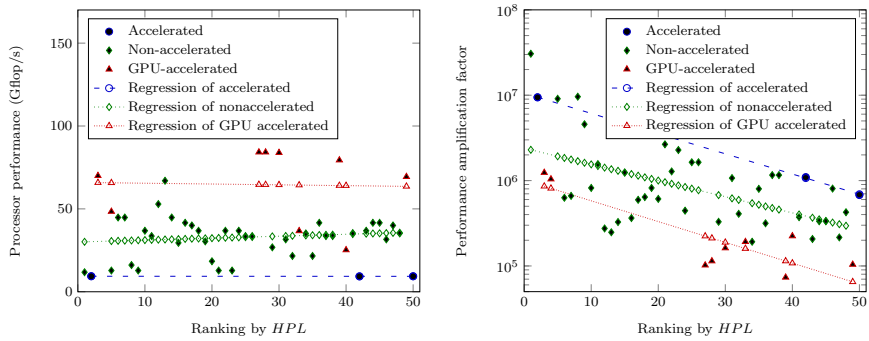
\begin{figure*}
	\maxsizebox{\textwidth}{!}
	{
%\MEfigure[]{
\begin{tabular}{rr}
\tikzset{mark options={mark size=2, line width=.5pt}}
		\begin{tikzpicture}
		\begin{axis}[%
		legend style={
			cells={anchor=west},
			legend pos={north west},
		},
		cycle list name={my color list},
		xmin=0, xmax=51,% x scale
		ymin=00, ymax=170, % y scale
		ylabel={Processor performance (Gflop/s)},
		xlabel={Ranking  by $HPL$} ,
		scatter/classes={%
			A={ mark=diamond*,  draw=webgreen},
			N={ mark=triangle*,  draw=webblue},
			G={ mark=square*,  draw=webred}
			}
		]
		\addplot[only marks, mark=*, draw=webblue,%
		scatter src=explicit symbolic]%
		table[meta=label] {
			x y label
			2	9.37	A
			42  9.25    A
			50  9.37    A
		};
		\addlegendentry{Accelerated}

		\addplot[only marks, mark=diamond*, draw=webgreen,%
		scatter src=explicit symbolic]%
		table[meta=label] {
			x y label
			1	11.78	N
			5  12.8   N
			6  44.8    N
			7  44.8    N
			8  16.1  N
			9  12.8  N
		   10  36.8  N
		   11  33.6  N
		   12  52.9  N
		   13  66.96 N
		   14  44.8  N
		   15  29.5  N
		   16  41.6  N
		   17  40.0  N
		   18  36.8  N
		   19  30.4  N
		   20  18.4  N
		   21  12.8  N
		   22  36.8  N
		   23  12.8  N
		   24  36.8  N
		   25  33.6  N
		   26  33.6  N
		   29  26.8  N
		   31  31.6  N
		   32  21.6  N
		   34  35.2  N
		   35  21.6  N
		   36  41.6  N
		   37  33.6  N
		   38  33.6  N
		   41  35.4  N
		   43  36.8  N
		   44  41.6  N
		   45  41.6  N 
		   46  31.6  N
		   47  40    N
		   48  35.2  N    	
		};
		\addlegendentry{Non-accelerated}
		
		\addplot[only marks, mark=triangle*, draw=webred,%
		scatter src=explicit symbolic]%
		table[meta=label] {
			x y label
			3	70.0	G
			5   48.4    G
		   27   84.2    G
		   28   84.2    G
		   30   83.9    G
		   33   36.7    G
		   39   79.4    G
		   40   25.2    G
		   49   69.4    G
		};
		\addlegendentry{GPU-accelerated}
		
		\addplot+[ mark=o,  draw=webblue] table[y={create col/linear regression={y=Y}},% mark=rectangle*,
		meta=label,    /pgf/number format/read comma as period
		] {
			x Y label
			2	9.37	A
			42  9.25    A
			50  9.37    A
		};
		\addlegendentry{Regression of accelerated}

		\addplot+[ mark=diamond,  draw=webgreen] table[y={create col/linear regression={y=Y}},% mark=rectangle*,
		meta=label,    /pgf/number format/read comma as period
		] {
			x Y label
			1	11.78	N
			5  12.8   N
			6  44.8    N
			7  44.8    N
			8  16.1  N
			9  12.8  N
		   10  36.8  N
		   11  33.6  N
		   12  52.9  N
		   13  66.96 N
		   14  44.8  N
		   15  29.5  N
		   16  41.6  N
		   17  40.0  N
		   18  36.8  N
		   19  30.4  N
		   20  18.4  N
		   21  12.8  N
		   22  36.8  N
		   23  12.8  N
		   24  36.8  N
		   25  33.6  N
		   26  33.6  N
		   29  26.8  N
		   31  31.6  N
		   32  21.6  N
		   34  35.2  N
		   35  21.6  N
		   36  41.6  N
		   37  33.6  N
		   38  33.6  N
		   41  35.4  N
		   43  36.8  N
		   44  41.6  N
		   45  41.6  N 
		   46  31.6  N
		   47  40    N
		   48  35.2  N    	
		};
		\addlegendentry{Regression of nonaccelerated}

		\addplot+[ mark=triangle,  draw=webred] table[y={create col/linear regression={y=Y}},% mark=rectangle*,
		meta=label,    /pgf/number format/read comma as period
		] {
			x Y label
			3	70.0	G
			5   48.4    G
		   27   84.2    G
		   28   84.2    G
		   30   83.9    G
		   33   36.7    G
		   39   79.4    G
		   40   25.2    G
		   49   69.4    G
		};
		\addlegendentry{Regression of GPU accelerated}
		
		\end{axis}
		\end{tikzpicture}
%	}

	&
\tikzset{mark options={mark size=2, line width=.5pt}}

		\begin{tikzpicture}
		\begin{axis}[%
		legend style={
			cells={anchor=west},
			legend pos={north east},
		},
		cycle list name={my color list},
		xmin=0, xmax=51,% x scale
		ymin=5e4, ymax=1e8, % y scale
		xlabel={Ranking  by $HPL$} ,
		ylabel={Performance amplification factor},
%		xmode=log,
		ymode=log,
		scatter/classes={%
			A={ mark=diamond*,  draw=webgreen},
			N={ mark=triangle*,  draw=webblue},
			G={ mark=square*,  draw=webred}
			}
		]
		\addplot[only marks,mark=*, draw=webblue,%
		scatter src=explicit symbolic]%
		table[meta=label] {
			x y label
			2	0.943e7	A
			42  0.110e7    A
			50  0.676e6    A
		};
		\addlegendentry{Accelerated}
		
		\addplot[only marks, mark=diamond*, draw=webgreen,%
		scatter src=explicit symbolic]%
		table[meta=label] {
			x y label
			1	0.306e8	N
			5  0.909e7   N
			6  0.629e6    N
			7  0.662e6    N
			8  0.961e7  N
			9  0.457e7  N
		   10  0.820e6  N
		   11  0.156e7  N
		   12  0.275e6  N
		   13  0.248e6 N
		   14  0.327e6  N
		   15  0.124e7  N
		   16  0.364e6  N
		   17  0.595e6  N
		   18  0.641e6  N
		   19  0.820e6  N
		   20  0.610e6  N
		   21  0.266e7  N
		   22  0.128e7  N
		   23  0.228e7  N
		   24  0.444e6  N
		   25  0.164e7  N
		   26  0.164e7  N
		   29  0.329e6  N
		   31  0.107e7  N
		   32  0.408e6  N
		   34  0.192e6  N
		   35  0.8e6  N
		   36  0.316e6  N
		   37  0.116e7  N
		   38  0.116e7  N
		   41  0.373e6  N
		   43  0.207e6  N
		   44  0.334e6  N
		   45  0.334e6  N 
		   46  0.806e6  N
		   47  0.216e6    N
		   48  0.426e6  N    	
		};
		\addlegendentry{Non-accelerated}
		
		\addplot[only marks, mark=triangle*, draw=webred,%
		scatter src=explicit symbolic]%
		table[meta=label] {
			x y label
			3	0.124e7    G
			4   0.104e7    G
		   27   0.102e6    G
		   28   0.114e6    G
		   30   0.162e6    G
		   33   0.192e6    G
		   39   0.735e5    G
		   40   0.224e6    G
		   49   0.104e6    G
		};
		\addlegendentry{GPU-accelerated}

		\addplot+[ mark=o,  draw=webblue] table[y={create col/linear regression={y=Y}},% mark=rectangle*,
		meta=label,    /pgf/number format/read comma as period
		] {
			x Y label
			2	0.943e7	A
			42  0.110e7    A
			50  0.676e6    A
		};
		\addlegendentry{Regression of accelerated}
		\addplot+[ mark=diamond,  draw=webgreen] table[y={create col/linear regression={y=Y}},% mark=rectangle*,
		meta=label,    /pgf/number format/read comma as period
		] {
			x Y label
			1	0.306e8	N
			5  0.909e7   N
			6  0.629e6    N
			7  0.662e6    N
			8  0.961e7  N
			9  0.457e7  N
		   10  0.820e6  N
		   11  0.156e7  N
		   12  0.275e6  N
		   13  0.248e6 N
		   14  0.327e6  N
		   15  0.124e7  N
		   16  0.364e6  N
		   17  0.595e6  N
		   18  0.641e6  N
		   19  0.820e6  N
		   20  0.610e6  N
		   21  0.266e7  N
		   22  0.128e7  N
		   23  0.228e7  N
		   24  0.444e6  N
		   25  0.164e7  N
		   26  0.164e7  N
		   29  0.329e6  N
		   31  0.107e7  N
		   32  0.408e6  N
		   34  0.192e6  N
		   35  0.8e6  N
		   36  0.316e6  N
		   37  0.116e7  N
		   38  0.116e7  N
		   41  0.373e6  N
		   43  0.207e6  N
		   44  0.334e6  N
		   45  0.334e6  N 
		   46  0.806e6  N
		   47  0.216e6    N
		   48  0.426e6  N    	
		};
		\addlegendentry{Regression of nonaccelerated}
		
		\addplot+[ mark=triangle,  draw=webred] table[y={create col/linear regression={y=Y}},% mark=rectangle*,
		meta=label,    /pgf/number format/read comma as period
		] {
			x Y label
			3	0.124e7    G
			4   0.104e7    G
		   27   0.102e6    G
		   28   0.114e6    G
		   30   0.162e6    G
		   33   0.192e6    G
		   39   0.735e5    G
		   40   0.224e6    G
		   49   0.104e6    G
		};
		\addlegendentry{Regression of GPU accelerated}
		\end{axis}
		\end{tikzpicture}

	\\
\end{tabular}
}
	\caption{Correlation of the efficiency and the performance amplification with the ranking, for different acceleration methods.}
	\label{fig:EfficiencyVSranking}
\end{figure*}

\subsection{Single-processor performance}\label{sec:SingleProcessorPerformance}

A common myth is that, as suggested by Equ.~(\ref{eq:PerformanceMax}), the trivial way to increase
the absolute performance of a supercomputer is to increase the single-processor
performance of its processors. Since the single processor performance has reached its limits,
some kind of accelerators are frequently used for this goal. Fig.~\ref{fig:EfficiencyVSranking} 
shows how utilizing accelerators influences ranking of supercomputers.

As the left side of the figure depicts,
the coprocessor accelerated cores show up the lowest performance; they really can benefit from acceleration\footnote{In the number of the total cores the number of coprocessors is included}.
 \gls{GPU} accelerated processors 
really increase performance of processors by a factor of 2-3, however this increased performance is about 40..70 times lower than the nominal performance of the  \gls{GPU} accelerator. This result confirms results of a former study where an average factor 2.5 was found~\cite{Lee:GPUvsCPU2010}. The effect is attributed to the considerable overhead~\cite{EfficiacyAPU:2011}, and it was demonstrated that with improving the transfer performance, the computing performance can be considerably enhanced. 
Indirectly, this research also proved that the operating principle itself (i.e. that the data must be transferred to and from the \gls{GPU} memory; and recall that \gls{GPU}s do not have cache memory) takes some extra time. In terms of Amdahl's law,
this transfer time contributes to the non-parallelizable fraction,
i.e. increases $(1-\alpha_{eff})$, i.e. decreases the achievable performance gain. See also Fig.~\ref{fig:PerformanceVSnoofprocessors}.

The right side of the figure discovers this effect. The performance amplification factor of the \gls{GPU} accelerated systems is about ten times worse than that of the coprocessor-accelerated processors and about 5 times worse than that of the the non-accelerated processors, i.e. the resulting efficiency is \textit{worse} than in the case of utilizing unaccelerated processors; this is a definite disadvance when \gls{GPU}s used in system with extremely large number of processors.
This makes at least questionable whether it is worth to utilize \gls{GPU}s 
in supercomputers. 

As the left figure shows, neither type of processors shows correlation between ranking of supercomputer and type of the acceleration. 
Essentially the same is confirmed by the right side of the figure:
the performance gain decreases with the ranking position:
to move the data form one memory to other takes time.

\subsection{Number of processors}

\begin{figure*}
%\MEfigure[]{
	\maxsizebox{\textwidth}{!}
	{
\begin{tabular}{rr}
\tikzset{mark options={mark size=2, line width=.5pt}}
\tikzset{mark options={mark size=2, line width=.5pt}}
		\begin{tikzpicture}
		\begin{axis}[%
		legend style={
			cells={anchor=west},
			legend pos={north east},
		},
		cycle list name={my color list},
		xmin=0, xmax=51,% x scale
		ymin=.03, ymax=11, % y scale
		ylabel={No of Processors/1e6},
		xlabel={Ranking  by $HPL$} ,
		ymode=log,
		scatter/classes={%
			A={ mark=diamond*,  draw=webgreen},
			N={ mark=triangle*,  draw=webblue},
			G={ mark=square*,  draw=webred}
			}
		]
		\addplot[only marks, mark=diamond*,  draw=webgreen,%
		scatter src=explicit symbolic]%
		table[meta=label] {
			x y label
			1	10.649600	A
			2   3.120000    A
			3   0.361760    A
			4   0.560640    A
			5   1.572864    A
			6   0.622336    A
			7   0.556104    A
			8   0.705024    A
			9   0.786432    A
		   10   0.301056    A
		   11  	0.241920    A	   
		   12  	0.241920    A	   
		   13   0.148716    A
		   14   0.241808    A
		   15	0.241108    A
		   16   0.231424    A
		   17   0.185088    A
		   18   0.185088    A
		   19   0.220800    A
		   20   0.522080    A
		   21   0.458752    A
		   22   0.144900    A
		   23   0.393216    A
		   24   0.145920    A
		   25   0.126468    A
		   26   0.126468    A
		   27   0.072800    A
		   28   0.072800    A
		   29   0.124200    A
		   30   0.072000    A
		   31   0.110160    A
		   32   0.225984    A
		   33   0.152692    A
		   34   0.092160    A
		   35   0.147456    A
		   36   0.086016    A
		   37   0.089856    A
		   38   0.089856    A
		   39   0.074520    A
		   40   0.186368    A
		   41   0.088992    A
		   42   0.194616    A
		   43   0.100064    A
		   44   0.069600    A
		   45   0.069600    A
		   46   0.082944    A
		   47   0.076032    A
		   48   0.072000    A
		   49   0.042688    A
		   50   0.174720    A
		};
		\addlegendentry{Data points}

		\addplot+[ mark=diamond,  draw=webgreen] table[y={create col/linear regression={y=Y}},% mark=rectangle*,
		meta=label,    /pgf/number format/read comma as period
		] {
			x Y label
			1	10.649600	A
			2   3.120000    A
			3   0.361760    A
			4   0.560640    A
			5   1.572864    A
			6   0.622336    A
			7   0.556104    A
			8   0.705024    A
			9   0.786432    A
		   10   0.301056    A
		   11  	0.241920    A	   
		   12  	0.241920    A	   
		   13   0.148716    A
		   14   0.241808    A
		   15	0.241108    A
		   16   0.231424    A
		   17   0.185088    A
		   18   0.185088    A
		   19   0.220800    A
		   20   0.522080    A
		   21   0.458752    A
		   22   0.144900    A
		   23   0.393216    A
		   24   0.145920    A
		   25   0.126468    A
		   26   0.126468    A
		   27   0.072800    A
		   28   0.072800    A
		   29   0.124200    A
		   30   0.072000    A
		   31   0.110160    A
		   32   0.225984    A
		   33   0.152692    A
		   34   0.092160    A
		   35   0.147456    A
		   36   0.086016    A
		   37   0.089856    A
		   38   0.089856    A
		   39   0.074520    A
		   40   0.186368    A
		   41   0.088992    A
		   42   0.194616    A
		   43   0.100064    A
		   44   0.069600    A
		   45   0.069600    A
		   46   0.082944    A
		   47   0.076032    A
		   48   0.072000    A
		   49   0.042688    A
		   50   0.174720    A
		};
		\addlegendentry{Regression Top50}
		
		\addplot+[ mark=*,  draw=webred, color=red] table[y={create col/linear regression={y=Y}},% mark=rectangle*,
		meta=label,    /pgf/number format/read comma as period
		] {
			x Y label
			1	10.649600	A
			2   3.120000    A
			3   0.361760    A
			4   0.560640    A
			5   1.572864    A
			6   0.622336    A
			7   0.556104    A
			8   0.705024    A
			9   0.786432    A
		   10   0.301056    A
		};
		\addlegendentry{Regression Top10}
		\end{axis}
		\end{tikzpicture}
%	}

	&
%	\maxsizebox{\columnwidth}{!}
%	{
\tikzset{mark options={mark size=2, line width=1pt}}
		\begin{tikzpicture}
		\begin{axis}[%
		legend style={
			cells={anchor=west},
			legend pos={north east},
		},
		cycle list name={my color list},
		xmin=0.05, xmax=12,% x scale
		ymin=2e-8, ymax=2e-5, % y scale
		xlabel={No of Processors/1e6} ,
		ylabel={(1-$\alpha_{eff}$  by $HPL$)},
		xmode=log,
		ymode=log,
		scatter/classes={%
			A={ mark=diamond*,  draw=webgreen},
			N={ mark=triangle*,  draw=webblue},
			G={ mark=square*,  draw=webred}
			}
		]
		\addplot[only marks,  mark=diamond*,  draw=webgreen,%
		scatter src=explicit symbolic]%
		table[meta=label] {
			x Y label
			10.649600 3.273e-8	A
			3.120000  1.991e-7  A
			0.361760  8.094E-07  A
			0.560640  9.656E-07 A
			1.572864  1.096E-07 A
            0.622336  1.590E-06 A
			0.556104  1.507E-06  A
			0.705024  1.040E-07  A
			0.786432  2.191E-07 A
		    0.301056  1.221E-06   A
		    0.241920  6.399E-07 A
		    0.241920  3.636E-06  A	   
		    0.148716  4.028E-06  A
		    0.241808  3.064E-06 A
		    0.241108  8.052E-07  A
		    0.231424  2.748E-06  A
		    0.185088  1.689E-06 A
		    0.185088  1.560E-06 A
		    0.220800  1.225E-06 A
		    0.522080  1.642E-06 A
		    0.458752  3.756E-07 A
		    0.144900  7.842E-07 A
		    0.393216  4.383E-07 A
		    0.145920  2.250E-06 A
		    0.126468  6.107E-07 A
		    0.126468  6.107E-07 A
		    0.072800  9.811E-06 A
		    0.072800  9.811E-06 A
		    0.124200  3.036E-06 A
		    0.072000  6.173E-06 A
		    0.110160  9.318E-07 A
		    0.225984  2.446E-06 A
		    0.152692  5.204E-06 A
		    0.092160  1.246E-06 A
		    0.147456  6.743E-07 A
		    0.086016  3.160E-06 A
		    0.089856  8.635E-07 A
		    0.089856  8.635E-07 A
		    0.074520  1.365E-05 A
		    0.186368  4.464E-06 A
		    0.088992  2.677E-06 A
		    0.194616  1.718E-06 A
		    0.100064  4.815E-06 A
		    0.069600  2.997E-06 A
		    0.069600  2.997E-06 A
		    0.082944  1.243E-06 A
		    0.076032  4.628E-06 A
		    0.072000  2.347E-06 A
		    0.042688  9.587E-06 A
		    0.174720  2.772E-06 A
		};
		\addlegendentry{Data points}
		
		\addplot+[ mark=diamond,  draw=webgreen] table[y={create col/linear regression={y=Y}},% mark=rectangle*,
		meta=label,    /pgf/number format/read comma as period
		] {
			x Y label
			10.649600 3.273e-8	A
			3.120000  1.991e-7  A
			0.361760  8.094E-07  A
			0.560640  9.656E-07 A
			1.572864  1.096E-07 A
            0.622336  1.590E-06 A
			0.556104  1.507E-06  A
			0.705024  1.040E-07  A
			0.786432  2.191E-07 A
		    0.301056  1.221E-06   A
		    0.241920  6.399E-07 A
		    0.241920  3.636E-06  A	   
		    0.148716  4.028E-06  A
		    0.241808  3.064E-06 A
		    0.241108  8.052E-07  A
		    0.231424  2.748E-06  A
		    0.185088  1.689E-06 A
		    0.185088  1.560E-06 A
		    0.220800  1.225E-06 A
		    0.522080  1.642E-06 A
		    0.458752  3.756E-07 A
		    0.144900  7.842E-07 A
		    0.393216  4.383E-07 A
		    0.145920  2.250E-06 A
		    0.126468  6.107E-07 A
		    0.126468  6.107E-07 A
		    0.072800  9.811E-06 A
		    0.072800  9.811E-06 A
		    0.124200  3.036E-06 A
		    0.072000  6.173E-06 A
		    0.110160  9.318E-07 A
		    0.225984  2.446E-06 A
		    0.152692  5.204E-06 A
		    0.092160  1.246E-06 A
		    0.147456  6.743E-07 A
		    0.086016  3.160E-06 A
		    0.089856  8.635E-07 A
		    0.089856  8.635E-07 A
		    0.074520  1.365E-05 A
		    0.186368  4.464E-06 A
		    0.088992  2.677E-06 A
		    0.194616  1.718E-06 A
		    0.100064  4.815E-06 A
		    0.069600  2.997E-06 A
		    0.069600  2.997E-06 A
		    0.082944  1.243E-06 A
		    0.076032  4.628E-06 A
		    0.072000  2.347E-06 A
		    0.042688  9.587E-06 A
		    0.174720  2.772E-06 A
		};
		\addlegendentry{Regression TOP50}

		\addplot+[ mark=*,  draw=webred] table[y={create col/linear regression={y=Y}},% mark=rectangle*,
		meta=label,    /pgf/number format/read comma as period
		] {
			x Y label
			10.649600 3.273e-8	A
			3.120000  1.991e-7  A
			0.361760  8.094E-07  A
			0.560640  9.656E-07 A
			1.572864  1.096E-07 A
            0.622336  1.590E-06 A
			0.556104  1.507E-06  A
			0.705024  1.040E-07  A
			0.786432  2.191E-07 A
		    0.301056  1.221E-06   A
		};
		\addlegendentry{Regression TOP10}
		
		\end{axis}
		\end{tikzpicture}
%	}

	\\
\end{tabular}
}
	\caption{Correlation of number of processors with ranking and effective parallelism with number of processors.}
	\label{fig:ProcNoVSranking}
\end{figure*}
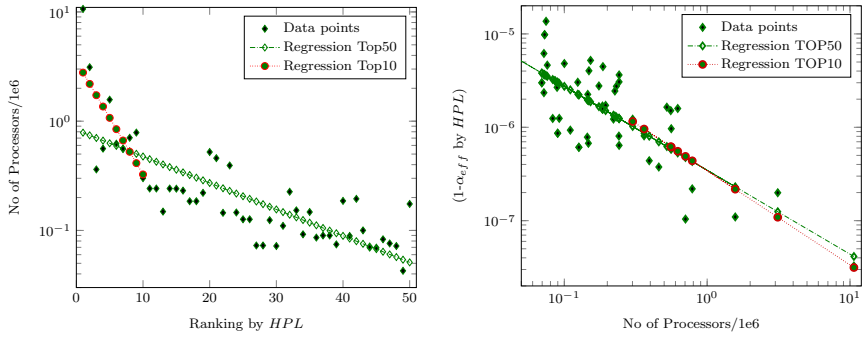

%}
%{Correlation of ranking and $\alpha_{eff}$, derived using  \gls{HPL} and \gls{HPCG}.}
%{fig:RankingVsAlpha}{}{}

Since the resulting performance depends both on the number of processors  and the effective parallelization,
both quantities are correlated in Fig.~\ref{fig:ProcNoVSranking}.
As expected, in TOP50 the higher the ranking position is,
the higher is the required number of processors in the configuration,
and as outlined above, the more processors, the lower $(1-\alpha_{eff})$
is required (provided that the same efficiency is targeted).

In TOP10, the slope of the regression line sharply changes
in the left figure, showing the strong competition for the better ranking position. 
Maybe this marks the cut line between the "race supercomputers" and "commodity supercomputers".
On the right figure, TOP10 data points provide the same slope as TOP50
data points, demonstrating that to produce a reasonable efficiency,
the increasing number of cores must be accompanied with a proper decrease in value of $(1-\alpha_{eff})$, as expected from Equ.~(\ref{eq:soverk}),
furthermore, that to achieve a good ranking a good value of $(1-\alpha_{eff})$ must be provided.

% % The performance amplification version
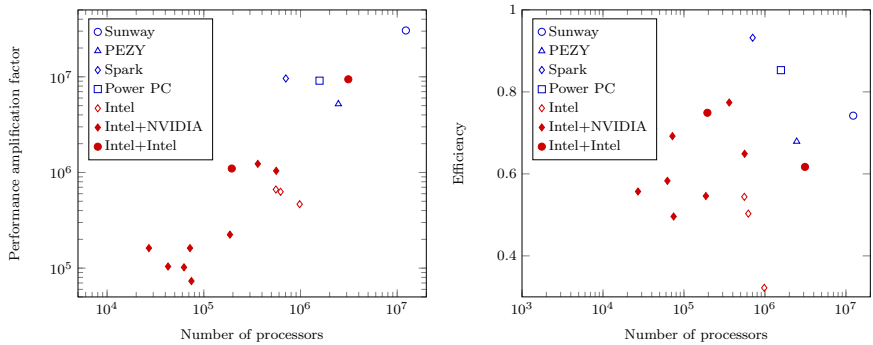
\begin{figure*}
	\maxsizebox{\textwidth}{!}
	{
%\MEfigure[]{
\begin{tabular}{rr}
\tikzset{mark options={mark size=2, line width=.5pt}}
		\begin{tikzpicture}
		\begin{axis}[%
		legend style={
			cells={anchor=west},
			legend pos={north west},
		},
%		cycle list name={my color list},
		ymin=5e4, ymax=5e7,% x scale
		xmin=5e3, xmax=2e7, % y scale
		ylabel={Performance amplification factor},
		xlabel={Number of processors},
		xmode=log,
		ymode=log,
		scatter/classes={%
			A={ mark=diamond*,  draw=webgreen},
			N={ mark=triangle*,  draw=webblue},
			G={ mark=square*,  draw=webred},
			C={ mark= square, draw =webred}
			}
		]
		
		\addplot[only marks,
		mark=o,draw=webblue,%
		scatter src=explicit symbolic]%
		table[meta=label] {
			x y label
			% #1
			12288000	0.305E08 C
		};
		\addlegendentry{Sunway}

				\addplot[only marks, mark=triangle,draw=webblue,%
		scatter src=explicit symbolic]%
		table[meta=label] {
			x y label
			% #4 Gyoukou
			2462640 0.520E07 PZ	
		};
		\addlegendentry{PEZY}
		
		\addplot[only marks, mark=diamond,draw=webblue,%
		scatter src=explicit symbolic]%
		table[meta=label] {
			x y label
			% #10 K
			705024 0.961E07 S	
		};
		\addlegendentry{Spark}

		\addplot[only marks, mark=square,draw=webblue,%
		scatter src=explicit symbolic]%
		table[meta=label] {
			x y label
			% #6 Sequuoia
			1572864 0.911E07  PW	
		};
		\addlegendentry{Power~PC}
		
		\addplot[only marks, mark=diamond,draw=webred,%
		scatter src=explicit symbolic]%
		table[meta=label] {
			x y label
			% #7 Trinity
			979968 0.466E06 I	
			% #8 Cori
			622336 0.628E06 I
			% #9 OakForest 
			556104 0.664E06 I					
		};
		\addlegendentry{Intel}

		\addplot[only marks, mark=diamond*,draw=webred,color=webred,%
		scatter src=explicit symbolic]%
		table[meta=label] {
			x y label
			  % Piz Daint, #3
			361760	0.12307E07 IN
			% Titan, #5
			560640 0.104E07 IN
			% #28,29 
			62400 0.102E06 IN
			% #31
			72000 0.162E06 IN
		    % #34
		    27056 0.162E06 IN
		    % #40
		    74520 0.733E05 IN
		    % #41
		    186368 0.224E06 IN
		    % #50 
			42688 0.104E06 IN		    			
		};
		\addlegendentry{Intel+NVIDIA}
		
		\addplot[only marks, mark=*,draw=webred, color=webred,%
		scatter src=explicit symbolic]%
		table[meta=label] {
			x y label
			% #2
			3120000 0.942E07 II			
		    % #43
		    194616 0.110E07 II		    		    			
		};
		\addlegendentry{Intel+Intel}

		\end{axis}
		\end{tikzpicture}
	&
		\begin{tikzpicture}
		\begin{axis}[%
		legend style={
			cells={anchor=west},
			legend pos={north west},
		},
%		cycle list name={my color list},
		ymin=0.3, ymax=1,% x scale
		xmin=1e3, xmax=2e7, % y scale
		ylabel={Efficiency},
		xlabel={Number of processors},
		xmode=log,
		]
		
		\addplot[only marks,
		mark=o,draw=webblue,%
		scatter src=explicit symbolic]%
		table[meta=label] {
			x y label
			% #1
			12288000	0.742 C
		};
		\addlegendentry{Sunway}

				\addplot[only marks, mark=triangle,draw=webblue,%
		scatter src=explicit symbolic]%
		table[meta=label] {
			x y label
			% #4 Gyoukou
			2462640 0.679 PZ	
		};
		\addlegendentry{PEZY}
		
		\addplot[only marks, mark=diamond,draw=webblue,%
		scatter src=explicit symbolic]%
		table[meta=label] {
			x y label
			% #10 K
			705024 0.932 S	
		};
		\addlegendentry{Spark}

		\addplot[only marks, mark=square,draw=webblue,%
		scatter src=explicit symbolic]%
		table[meta=label] {
			x y label
			% #6 Sequuoia
			1572864 0.853  PW	
		};
		\addlegendentry{Power~PC}
		
		\addplot[only marks, mark=diamond,draw=webred,%
		scatter src=explicit symbolic]%
		table[meta=label] {
			x y label
			% #7 Trinity
			979968 0.322 I	
			% #8 Cori
			622336 0.503 I
			% #9 OakForest 
			556104 0.544 I					
		};
		\addlegendentry{Intel}

		\addplot[only marks, mark=diamond*,draw=webred,color=webred,%
		scatter src=explicit symbolic]%
		table[meta=label] {
			x y label
			  % Piz Daint, #3
			361760	0.774 IN
			% Titan, #5
			560640 0.649 IN
			% #28,29 
			62400 0.583 IN
			% #31
			72000 0.692 IN
		    % #34
		    27056 0.557 IN
		    % #40
		    74520 0.496 IN
		    % #41
		    186368 0.546 IN
		    % #50 
			42688 0.710 IN		    			
		};
		\addlegendentry{Intel+NVIDIA}
		
		\addplot[only marks, mark=*,draw=webred, color=webred,%
		scatter src=explicit symbolic]%
		table[meta=label] {
			x y label
			% #2
			3120000 0.617 II			
		    % #43
		    194616 0.749 II		    		    			
		};
		\addlegendentry{Intel+Intel}

		\end{axis}
		\end{tikzpicture}
	\\
\end{tabular}
}
	\caption{Correlation of the performance amplification and the efficiency with the number of processors, for some Intel based systems, with and without acceleration. For comparison data for some other processors are also depicted.}
	\label{fig:PerformanceVSnoofprocessors}
\end{figure*}

The effect of acceleration, discussed in section~\ref{sec:SingleProcessorPerformance}, can also be scrutinized under more clean conditions, in function of the number of the cores rather than in function of the payload performance. To make further cleanup, only data about processors from the same manufacturer are depicted in Fig.~\ref{fig:PerformanceVSnoofprocessors}. 
As shown, \gls{GPU} acceleration results in both rather wrong performance amplification parameters and efficiency, even at processor numbers below $10^5$.  In other words: deploying \gls{GPU}-accelerated cores in supercomputers having millions of processors is a rather expensive way to make supercomputer performance \textit{worse}.

\subsection{Architectural solution}

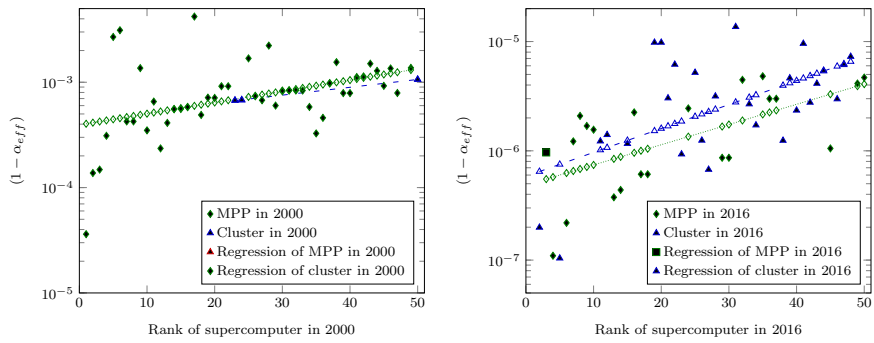
\begin{figure*}
	\maxsizebox{\textwidth}{!}
	{
\begin{tabular}{rr}
		\begin{tikzpicture}
		\begin{axis}[%
		legend style={
			cells={anchor=west},
			legend pos={south east},
		},
		cycle list name={my color list},
		xmin=0, xmax=51,% x scale
		ymode=log,
		ymin=1e-5, ymax=5e-3, % y scale
		xlabel=Rank of supercomputer in 2000,
		ylabel=$(1-\alpha_{eff})$ ,
		scatter/classes={%
			MPP={ mark=diamond*,  draw=webgreen}, Cluster={ mark=triangle*, draw=webblue}, 
			%			MPPR={ mark=square*,  draw=webgreen}, ClusterR={ mark=triangle*, draw=webblue},
			SCMCS={mark=triangle*,draw=webred} }
		]
		\addplot[scatter,only marks, %
		scatter src=explicit symbolic]%
		table[meta=label,    /pgf/number format/read comma as period
		] {
			x y label
			1	3,614E-05	MPP
			2	1,375E-04	MPP
			3	1,482E-04	MPP
			4	3,103E-04	MPP
			5	2,690E-03	MPP
			6	3,117E-03	MPP
			7	4,247E-04	MPP
			8	4,247E-04	MPP
			9	1,362E-03	MPP
			10	3,493E-04	MPP
			11	6,552E-04	MPP
			12	2,355E-04	MPP
			13	4,112E-04	MPP
			14	5,575E-04	MPP
			15	5,575E-04	MPP
			16	5,811E-04	MPP
			17	4,201E-03	MPP
			18	4,894E-04	MPP
			19	7,143E-04	MPP
			20	7,102E-04	MPP
			21	9,167E-04	MPP
			22	9,167E-04	MPP
			25	1,685E-03	MPP
			26	7,362E-04	MPP
			27	6,746E-04	MPP
			28	2,227E-03	MPP
			29	6,005E-04	MPP
			30	8,343E-04	MPP
			31	8,343E-04	MPP
			32	8,343E-04	MPP
			33	8,343E-04	MPP
			34	5,828E-04	MPP
			35	3,267E-04	MPP
			36	4,592E-04	MPP
			37	9,823E-04	MPP
			38	1,551E-03	MPP
			39	7,889E-04	MPP
			40	7,889E-04	MPP
			41	1,120E-03	MPP
			42	1,136E-03	MPP
			43	1,500E-03	MPP
			44	1,282E-03	MPP
			45	9,241E-04	MPP
			46	1,352E-03	MPP
			47	7,905E-04	MPP
			49	1,369E-03	MPP
		};
		\addlegendentry{MPP in 2000}
		
		\addplot[scatter,only marks, mark=triangle*,
		scatter src=explicit symbolic]%
		table[meta=label, /pgf/number format/read comma as period] {
			x y label
			23	6,749E-04	Cluster
			24	6,749E-04	Cluster
			50	1,063E-03	Cluster
		};
		\addlegendentry{Cluster in 2000}
		
		\addplot+[ mark=diamond,  draw=webgreen] table[y={create col/linear regression={y=Y}},% mark=rectangle*,
		meta=label,    /pgf/number format/read comma as period
		] {
			x Y label
			1	3,614E-05	MPPR
			2	1,375E-04	MPPR
			3	1,482E-04	MPP
			4	3,103E-04	MPP
			5	2,690E-03	MPP
			6	3,117E-03	MPP
			7	4,247E-04	MPP
			8	4,247E-04	MPP
			9	1,362E-03	MPP
			10	3,493E-04	MPP
			11	6,552E-04	MPP
			12	2,355E-04	MPP
			13	4,112E-04	MPP
			14	5,575E-04	MPP
			15	5,575E-04	MPP
			16	5,811E-04	MPP
			17	4,201E-03	MPP
			18	4,894E-04	MPP
			19	7,143E-04	MPP
			20	7,102E-04	MPP
			21	9,167E-04	MPP
			22	9,167E-04	MPP
			25	1,685E-03	MPP
			26	7,362E-04	MPP
			27	6,746E-04	MPP
			28	2,227E-03	MPP
			29	6,005E-04	MPP
			30	8,343E-04	MPP
			31	8,343E-04	MPP
			32	8,343E-04	MPP
			33	8,343E-04	MPP
			34	5,828E-04	MPP
			35	3,267E-04	MPP
			36	4,592E-04	MPP
			37	9,823E-04	MPP
			38	1,551E-03	MPP
			39	7,889E-04	MPP
			40	7,889E-04	MPP
			41	1,120E-03	MPP
			42	1,136E-03	MPP
			43	1,500E-03	MPP
			44	1,282E-03	MPP
			45	9,241E-04	MPP
			46	1,352E-03	MPP
			47	7,905E-04	MPP
			49	1,369E-03	MPPR
		};
		\addlegendentry{Regression of MPP in 2000}
		
		\addplot+[ mark=triangle, draw=webblue] table[y={create col/linear regression={y=Y}}, mark=diamond*,
		meta=label,    /pgf/number format/read comma as period
		] {
			x Y label
			23	6,749E-04	Cluster
			24	6,749E-04	Cluster
			50	1,063E-03	Cluster
		};
		\addlegendentry{Regression of cluster in 2000}
		\end{axis}
		\end{tikzpicture}
	&
		\begin{tikzpicture}
		\begin{axis}[%
		legend style={
			cells={anchor=west},
			legend pos={south east},
		},
		cycle list name={my color list},
		xmin=0, xmax=51,% x scale
		ymode=log,
		ymin=5e-8, ymax=2e-5, % y scale
		xlabel=Rank of supercomputer in 2016,
		ylabel=$(1-\alpha_{eff})$,
		scatter/classes={%
			MPP={ mark=diamond*,  draw=webgreen}, Cluster={ mark=triangle*, draw=webblue}, 
			MPPR={ mark=square*,  draw=webgreen}, ClusterR={ mark=triangle*, draw=webblue},
			SCMCS={mark=triangle*,draw=webred} }
		]
		\addplot[scatter,only marks,%
		scatter src=explicit symbolic]%
		table[meta=label] {
			x y label
			3	9.656E-07	MPPR
			4	1.096E-07	MPP
			6	2.191E-07	MPP
			7	1.221E-06	MPP
			8	2.087E-06	MPP
			9	1.689E-06	MPP
			10	1.560E-06	MPP
			13	3.756E-07	MPP
			14	4.383E-07	MPP
			16	2.250E-06	MPP
			17	6.107E-07	MPP
			18	6.107E-07	MPP
			24	2.446E-06	MPP
			29	8.635E-07	MPP
			30	8.635E-07	MPP
			32	4.464E-06	MPP
			35	4.815E-06	MPP
			36	2.997E-06	MPP
			37	2.997E-06	MPP
			45	1.052E-06	MPP
			49	4.131E-06	MPP
			50	4.682E-06	MPP
		};
		\addlegendentry{MPP in 2016}
		
		\addplot[scatter,only marks,%
		scatter src=explicit symbolic]%
		table[meta=label] {
			x y label
			2	1.991E-07	ClusterR
			5	1.040E-07	Cluster
			11	1.225E-06	Cluster
			12	1.402E-06	Cluster
			15	1.163E-06	Cluster
			19	9.811E-06	Cluster
			20	9.811E-06	Cluster
			21	3.036E-06	Cluster
			22	6.173E-06	Cluster
			23	9.318E-07	Cluster
			25	5.204E-06	Cluster
			26	1.246E-06	Cluster
			27	6.743E-07	Cluster
			28	3.160E-06	Cluster
			31	1.365E-05	Cluster
			33	2.677E-06	Cluster
			34	1.718E-06	Cluster
			38	1.243E-06	Cluster
			39	4.628E-06	Cluster
			40	2.347E-06	Cluster
			41	9.587E-06	Cluster
			42	2.772E-06	Cluster
			43	4.132E-06	Cluster
			44	5.438E-06	Cluster
			46	2.976E-06	Cluster
			47	6.123E-06	Cluster
			48	7.291E-06	Cluster
		};
		\addlegendentry{Cluster in 2016}
		
		\addplot+[ mark=diamond,  draw=webgreen] table[y={create col/linear regression={y=Y}}, %mark=rectangle*,
		meta=label,    /pgf/number format/read comma as period
		]
		{
			x Y label
			3	9.656E-07	MPP
			4	1.096E-07	MPP
			6	2.191E-07	MPP
			7	1.221E-06	MPP
			8	2.087E-06	MPP
			9	1.689E-06	MPP
			10	1.560E-06	MPP
			13	3.756E-07	MPP
			14	4.383E-07	MPP
			16	2.250E-06	MPP
			17	6.107E-07	MPP
			18	6.107E-07	MPP
			24	2.446E-06	MPP
			29	8.635E-07	MPP
			30	8.635E-07	MPP
			32	4.464E-06	MPP
			35	4.815E-06	MPP
			36	2.997E-06	MPP
			37	2.997E-06	MPP
			45	1.052E-06	MPP
			49	4.131E-06	MPP
			50	4.682E-06	MPP
		};
		\addlegendentry{Regression of MPP in 2016}
		
		\addplot+[ mark=triangle, draw=webblue] table[y={create col/linear regression={y=Y}}, %mark=o*,
		meta=label,    /pgf/number format/read comma as period
		]
		{
			x Y label
			2	1.991E-07	Cluster
			5	1.040E-07	Cluster
			11	1.225E-06	Cluster
			12	1.402E-06	Cluster
			15	1.163E-06	Cluster
			19	9.811E-06	Cluster
			20	9.811E-06	Cluster
			21	3.036E-06	Cluster
			22	6.173E-06	Cluster
			23	9.318E-07	Cluster
			25	5.204E-06	Cluster
			26	1.246E-06	Cluster
			27	6.743E-07	Cluster
			28	3.160E-06	Cluster
			31	1.365E-05	Cluster
			33	2.677E-06	Cluster
			34	1.718E-06	Cluster
			38	1.243E-06	Cluster
			39	4.628E-06	Cluster
			40	2.347E-06	Cluster
			41	9.587E-06	Cluster
			42	2.772E-06	Cluster
			43	4.132E-06	Cluster
			44	5.438E-06	Cluster
			46	2.976E-06	Cluster
			47	6.123E-06	Cluster
			48	7.291E-06	Cluster
		};
		\addlegendentry{Regression of cluster in 2016}
		\end{axis}
		\end{tikzpicture}
	\\
\end{tabular}
}
	\caption{Dependence of $(1-\alpha_{eff})$ on the architectural solution of supercomputer in 2000 and 2016.  Data derived using the \gls{HPL} benchmark.}
	\label{fig:SCarchitecture}
\end{figure*}

Another common myth is that the internal interconnection method can considerably enhance the effective parallelism. 
As shown in Fig.~\ref{fig:SCarchitecture}, with time the composition of the type of the architectural solutions as well as the value of 
parallelization efficiency have considerably changed. However, in neither time the architectural solution caused significant difference compared to the other one; the slope is the same for both solutions, in both years. This means that the internal connection bandwidth is not a real bottleneck in improving performance. At the same time, $(1-\alpha_{eff})$ has improved independently and considerably.

\subsection{Benchmarking}

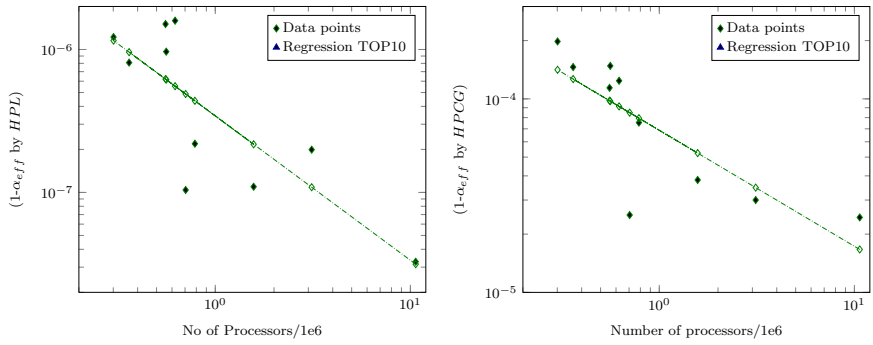
\begin{figure*}
	\maxsizebox{\textwidth}{!}
	{
\begin{tabular}{rr}
\tikzset{mark options={mark size=2, line width=.5pt}}
		\begin{tikzpicture}
		\begin{axis}[%
		legend style={
			cells={anchor=west},
			legend pos={north east},
		},
		cycle list name={my color list},
%		xmin=1e-8, xmax=5e-6,% x scale
		xmin=.2, xmax=12,% x scale
		ymin=2e-8, ymax=2e-6, % y scale
		ylabel={(1-$\alpha_{eff}$  by $HPL$)},
		xlabel={No of Processors/1e6},
		xmode=log,
		ymode=log,
		scatter/classes={%
			A={ mark=diamond*,  draw=webgreen},
			N={ mark=triangle*,  draw=webblue},
			G={ mark=square*,  draw=webred}
			}
		]
		\addplot[scatter,only marks,%
		scatter src=explicit symbolic]%
		table[meta=label] {
			x y label
		10.649600 3.273E-08 A
		3.120000  1.991E-07 A
		0.361760  8.094E-07 A
		0.560640  9.656E-07 A
		1.572864  1.096E-07 A
		0.622336  1.590E-06 A
		0.556104  1.507E-06 A
		0.705024  1.040E-07 A
		0.786432  2.191E-07 A
		0.301056  1.221E-06 A
		};
%		\addplot[scatter,only marks,%
%		scatter src=explicit symbolic]%
%		table[meta=label] {
%			x y label
%		3.273E-08 2.44E-05 A
%		1.991E-07 3.00E-05 A
%		8.094E-07 1.46E-04 A
%		9.656E-07 1.48E-04 A
%		1.096E-07 3.81E-05 A
%		1.590E-06 1.24E-04 A
%		1.507E-06 1.14E-04 A
%		1.040E-07 2.51E-05 A
%		2.191E-07 7.54E-05 A
%		1.221E-06 1.98E-04 A
%		};
		\addlegendentry{Data points}

		\addplot+[ mark=diamond,  draw=webgreen] table[y={create col/linear regression={y=Y}},% mark=rectangle*,
		meta=label,    /pgf/number format/read comma as period
		] {
			x Y label
		10.649600 3.273E-08 A
		3.120000  1.991E-07 A
		0.361760  8.094E-07 A
		0.560640  9.656E-07 A
		1.572864  1.096E-07 A
		0.622336  1.590E-06 A
		0.556104  1.507E-06 A
		0.705024  1.040E-07 A
		0.786432  2.191E-07 A
		0.301056  1.221E-06 A
		};
		\addlegendentry{Regression TOP10}
		
		\end{axis}
		\end{tikzpicture}
%	}

	&
		\begin{tikzpicture}
		\begin{axis}[%
		legend style={
			cells={anchor=west},
			legend pos={north east},
		},
		cycle list name={my color list},
		xmin=.2, xmax=12,% x scale
		ymin=1e-5, ymax=3e-4, % y scale
		ylabel={(1-$\alpha_{eff}$  by $HPCG$)},
		xlabel={Number of processors/1e6},
		xmode=log,
		ymode=log,
		scatter/classes={%
			A={ mark=diamond*,  draw=webgreen},
			N={ mark=triangle*,  draw=webblue},
			G={ mark=square*,  draw=webred}
			}
		]
		\addplot[scatter,only marks,%
		scatter src=explicit symbolic]%
		table[meta=label] {
			x y label
		10.649600 2.44E-05 A
		3.120000  3.00E-05 A
		0.361760  1.46E-04 A
		0.560640  1.48E-04 A
		1.572864  3.81E-05 A
		0.622336  1.24E-04 A
		0.556104  1.14E-04 A
		0.705024  2.51E-05 A
		0.786432  7.54E-05 A
		0.301056  1.98E-04 A
		};
		\addlegendentry{Data points}

		\addplot+[ mark=diamond,  draw=webgreen] table[y={create col/linear regression={y=Y}},% mark=rectangle*,
		meta=label,    /pgf/number format/read comma as period
		] {
			x Y label
		10.649600 2.44E-05 A
		3.120000  3.00E-05 A
		0.361760  1.46E-04 A
		0.560640  1.48E-04 A
		1.572864  3.81E-05 A
		0.622336  1.24E-04 A
		0.556104  1.14E-04 A
		0.705024  2.51E-05 A
		0.786432  7.54E-05 A
		0.301056  1.98E-04 A
		};
		\addlegendentry{Regression TOP10}
		
		\end{axis}
		\end{tikzpicture}
%	}

	\\
\end{tabular}
}
	\caption{Correlation of $(1-\alpha_{eff}^{HPL})$ and  $(1-\alpha_{eff}^{HPCG})$ with the number of processors.}
	\label{fig:Alpha_HPLvsHPCG}
\end{figure*}

According to the model, the \gls{SW} (including benchmark programs)
also contributes to the measured $(1-\alpha_{eff})$,
and its contribution is different for the different programs.
Fortunately, since the beginnings the same benchmark program \gls{HPL}
is used to qualify supercomputers.
\gls{HPL} contributes only a low amount of overhead activity,
so it can be used as the best estimator for describing the \gls{HW}+\gls{OS}
environment of a supercomputer.
Unfortunately, most real-life applications have much higher \gls{SW} contribution,
so recently benchmark \gls{HPCG} has been suggested to imitate their behavior.
Fig.~\ref{fig:Alpha_HPLvsHPCG} shows how $(1-\alpha_{eff})$ correlates
with number of processing units, for the two mentioned benchmark programs. The behavior is quite similar on the left and right figures, but 
the value differs by about two orders of magnitude.
Because of this, it can be safely stated that \gls{HPCG} 
measures the behavior of the program on the architecture
rather than the architecture ($\alpha_{eff}^{HW+OS}$) itself.
Notice also, how the relative $\alpha_{eff}$ measured values change between the two benchmarks.

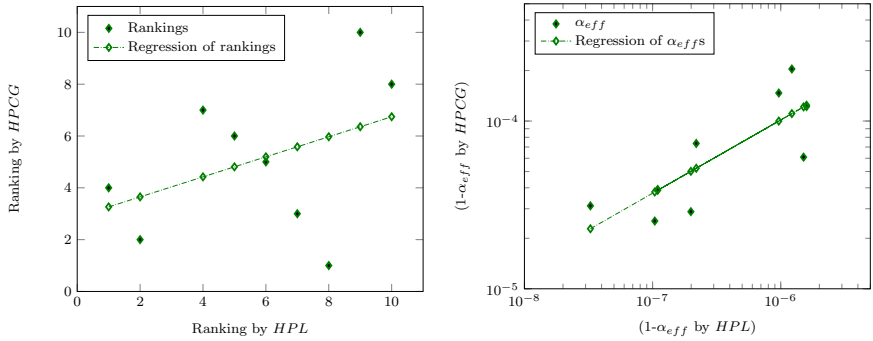
\begin{figure*}
	\maxsizebox{\textwidth}{!}
	{
\begin{tabular}{rr}
\tikzset{mark options={mark size=2, line width=1pt}}
		\begin{tikzpicture}
		\begin{axis}[%
		legend style={
			cells={anchor=west},
			legend pos={north west},
		},
		cycle list name={my color list},
		xmin=0, xmax=11,% x scale
		ymin=0, ymax=11, % y scale
		xlabel={Ranking  by $HPL$},
		ylabel={Ranking  by $HPCG$} ,
		scatter/classes={%
			MPP={ mark=diamond*,  draw=webgreen}}
		]
		\addplot[scatter,only marks,%
		scatter src=explicit symbolic]%
		table[meta=label] {
			x y label
			1	4	MPP
			2	2	MPP
			4   7   MPP
			5   6   MPP
			6   5   MPP
			7   3   MPP
			8   1   MPP
			9   10  MPP
		   10   8   MPP
		};
		\addlegendentry{Rankings}

		\addplot+[ mark=diamond,  draw=webgreen] table[y={create col/linear regression={y=Y}},% mark=rectangle*,
		meta=label,    /pgf/number format/read comma as period
		] {
			x Y label
			1	4	MPP
			2	2	MPP
			4   7   MPP
			5   6   MPP
			6   5   MPP
			7   3   MPP
			8   1   MPP
			9   10  MPP
		   10   8   MPP
		};
		\addlegendentry{Regression of rankings}
		
		\end{axis}
		\end{tikzpicture}
%	}

	&
\tikzset{mark options={mark size=2, line width=1pt}}
		\begin{tikzpicture}
		\begin{axis}[%
		legend style={
			cells={anchor=west},
			legend pos={north west},
		},
		cycle list name={my color list},
		xmin=1e-8, xmax=5e-6,% x scale
		ymin=1e-5, ymax=5e-4, % y scale
		xlabel={(1-$\alpha_{eff}$  by $HPL$)},
		ylabel={(1-$\alpha_{eff}$  by $HPCG$)},
		xmode=log,
		ymode=log,
		scatter/classes={%
			MPP={ mark=diamond*,  draw=webgreen}}
		]
		\addplot[scatter,only marks,
		scatter src=explicit symbolic]%
		table[meta=label] {
			x y label
			3.273E-08	3.121e-5	MPP
			1.991E-07	2.882e-5	MPP
			9.656E-07   1.469e-4   MPP
			1.096E-07   3.910e-5   MPP
			1.590E-06   1.220e-4   MPP
			1.507E-06   6.092e-5   MPP
			1.040E-07   2.534e-5   MPP
			2.191E-07   7.353e-5  MPP
		   1.221E-06   2.043e-4   MPP
		};
		\addlegendentry{$\alpha_{eff}$}

		\addplot+[ mark=diamond,  draw=webgreen] table[y={create col/linear regression={y=Y}},% mark=rectangle*,
		meta=label,    /pgf/number format/read comma as period
		] {
			x Y label
			3.273E-08	3.121e-5	MPP
			1.991E-07	2.882e-5	MPP
			9.656E-07   1.469e-4   MPP
			1.096E-07   3.910e-5   MPP
			1.590E-06   1.220e-4   MPP
			1.507E-06   6.092e-5   MPP
			1.040E-07   2.534e-5   MPP
			2.191E-07   7.353e-5  MPP
		   1.221E-06   2.043e-4   MPP
		};
		\addlegendentry{Regression of $\alpha_{eff}$s}
		
		\end{axis}
		\end{tikzpicture}
%	}

	\\
\end{tabular}
}
	\caption{Correlation of ranking and $\alpha_{eff}$, derived using  \gls{HPL} and \gls{HPCG}.}
	\label{fig:RankingVsAlpha}
\end{figure*}
%}
%{Correlation of ranking and $\alpha_{eff}$, derived using  \gls{HPL} and \gls{HPCG}.}
%{fig:RankingVsAlpha}{}{}
\subsection{Ranking}
For ranking, different merits can be used. 
One possible approach is to measure $R_{Max}$, using benchmarks either \gls{HPL} or \gls{HPG}.
Of course, these two measurements lead to different rankings.
Another possible approach is to rank by $\alpha_{eff}$,
measured with either of the two benchmarks.
Fig.~\ref{fig:RankingVsAlpha} compares how these two measurements correlate with each other.
Data points on the left figure show no correlation,
strongly supporting the statement that \gls{HPL} measures 
the architecture, \gls{HPCG} measures the \gls{SW} contribution, and so they are not correlated at all.
In contrast, the two $(1-\alpha_{eff})$ values strongly correlate, although the dominating contribution changes the order of magnitude on the two axes.

\subsection{Efficiency}

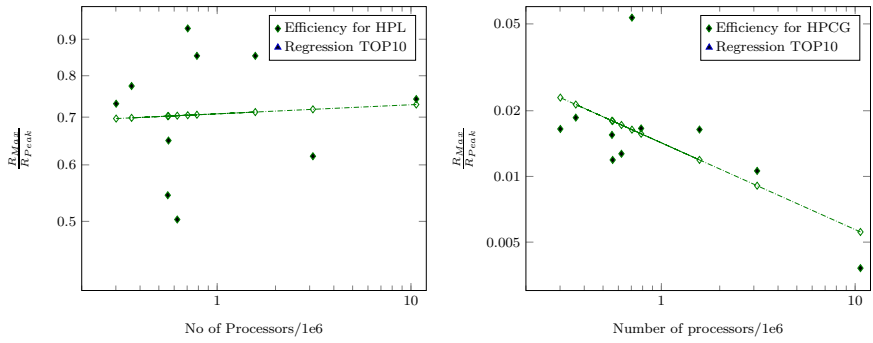
\begin{figure*}
	\maxsizebox{\textwidth}{!}
	{
\begin{tabular}{rr}
\tikzset{mark options={mark size=2, line width=.5pt}}
		\begin{tikzpicture}
		\begin{axis}[%
		legend style={
			cells={anchor=west},
			legend pos={north east},
		},
		cycle list name={my color list},
%		xmin=1e-8, xmax=5e-6,% x scale
		xmin=.2, xmax=12,% x scale
		ymin=.4, ymax=1, % y scale
		ylabel={$\frac{R_{Max}}{R_{Peak}}$},
		xlabel={No of Processors/1e6},
		xmode=log,
		ymode=log,
		ytick={.5,.6,.7,.8,.9},
        log ticks with fixed point,
		scatter/classes={%
			A={ mark=diamond*,  draw=webgreen},
			N={ mark=triangle*,  draw=webblue},
			G={ mark=square*,  draw=webred}
			}
		]
		\addplot[scatter,only marks,%
		scatter src=explicit symbolic]%
		table[meta=label] {
			x y label
		10.649600 0.742 A
		3.120000  0.617 A
		0.361760  0.774 A
		0.560640  0.649 A
		1.572864  0.853 A
		0.622336  0.503 A
		0.556104  0.544 A
		0.705024  0.932 A
		0.786432  0.853 A
		0.301056  0.731 A
		};
		\addlegendentry{Efficiency for HPL}

		\addplot+[ mark=diamond,  draw=webgreen] table[y={create col/linear regression={y=Y}},% mark=rectangle*,
		meta=label,    /pgf/number format/read comma as period
		] {
			x Y label
		10.649600 0.742 A
		3.120000  0.617 A
		0.361760  0.774 A
		0.560640  0.649 A
		1.572864  0.853 A
		0.622336  0.503 A
		0.556104  0.544 A
		0.705024  0.932 A
		0.786432  0.853 A
		0.301056  0.731 A
		};
		\addlegendentry{Regression TOP10}
		
		\end{axis}
		\end{tikzpicture}
%	}

	&
		\begin{tikzpicture}
		\begin{axis}[%
		legend style={
			cells={anchor=west},
			legend pos={north east},
		},
		cycle list name={my color list},
		xmin=.2, xmax=12,% x scale
		ymin=3e-3, ymax=6e-2, % y scale
		ylabel={$\frac{R_{Max}}{R_{Peak}}$},
		xlabel={Number of processors/1e6},
		xmode=log,
		ymode=log,
		ytick={5e-3,1e-2,2e-2,5e-2},
        log ticks with fixed point,
		scatter/classes={%
			A={ mark=diamond*,  draw=webgreen},
			N={ mark=triangle*,  draw=webblue},
			G={ mark=square*,  draw=webred}
			}
		]
		\addplot[scatter,only marks,%
		scatter src=explicit symbolic]%
		table[meta=label] {
			x y label
		10.649600 0.0038 A
		3.120000  0.0106 A
		0.361760  0.0186 A
		0.560640  0.0119 A
		1.572864  0.0164 A
		0.622336  0.0127 A
		0.556104  0.0155 A
		0.705024  0.0534 A
		0.786432  0.0166 A
		0.301056  0.0165 A
		};
		\addlegendentry{Efficiency for HPCG}
				
		\addplot+[ mark=diamond,  draw=webgreen] table[y={create col/linear regression={y=Y}},% mark=rectangle*,
		meta=label,    /pgf/number format/read comma as period
		] {
			x Y label
		10.649600 0.0038 A
		3.120000  0.0106 A
		0.361760  0.0186 A
		0.560640  0.0119 A
		1.572864  0.0164 A
		0.622336  0.0127 A
		0.556104  0.0155 A
		0.705024  0.0534 A
		0.786432  0.0166 A
		0.301056  0.0165 A
		};
		\addlegendentry{Regression TOP10}
		
		\end{axis}
		\end{tikzpicture}
%	}

	\\
\end{tabular}
}
	\caption{Correlation of efficiency with the number of processors, for the TOP10 supercomputers in 2017. Left: results for benchmark \gls{HPL}. Right: results for benchmark \gls{HPCG}. }
	\label{fig:EfficiencyVSProcNo}
\end{figure*}

Although $\frac{R_{Max}}{R_{Peak}}$ measured with benchmark \gls{HPL} is an important feature of the \gls{HW}+\gls{OS} assembly,
it is a reliable merit only when $\alpha_{eff}^{SW}$ is less than $\alpha_{eff}^{HW+OS}$.
As long as \gls{HPL} is used to rank supercomputers, architects keep efficiency around 
0.73; although in the case of Taihulight~\cite{FuSunwaySystem2016}
 (because of the extremely large number of processors) it is only possible through using special \gls{HW} units \gls{MPE}~\cite{CooperativeComputing2015}.
In the case of real-life programs, however, $\alpha_{eff}^{SW}$ is about two orders of magnitude higher
than $\alpha_{eff}^{HW+OS}$ (see Fig.~\ref{fig:Alpha_HPLvsHPCG}), so in that case the efficiency 
steeply decreases as the number of processors increases, see the right side of Fig.~\ref{fig:EfficiencyVSProcNo}.
Notice that the measured efficiency of Taihulight changes
drastically: utilizing \gls{MPE}s decreases $\alpha_{eff}^{HW+OS}$ which is considerable in the case of benchmark \gls{HPL},
but in the case of \gls{HPCG} $\alpha_{eff}^{SW}$ dominates, so the effect of \gls{MPE}s are negligible. 

It is important to notice, that $\alpha_{eff}$ sensitively changes with number of processors (see Fig.~\ref{fig:Alpha_HPLvsHPCG} ), while efficiency does not. 
This mostly follows from the fact that supercomputers are ranked based on benchmark \gls{HPL}.
When changing to \gls{HPCG}, the ranking -- and accordingly the 
direction of development -- will change, having effect also on these parameters.

\section{Future of supercomputers}\label{sec:future}

The race for achieving Eflop/s performance is continuing.
From the presently existing implementations some conclusions 
can be already drawn. From Fig.~3 %\ref{fig:PerformanceVSranking}
one can conclude for the near future  an optimistic single processor performance $P$ of value 50 Gflop/s. From Equ.~(\ref{eq:PerformanceMax}) one shall conclude that for achieving 1 Eflop/s payload performance $(1-\alpha)$ of value
$5\times 10^{-8}$ effective parallelization should be achieved.
Compare the values to the case of Taihulight:
$P=11.8$ Gflop/s and $(1-\alpha_{eff})= 3.3\times 10^{-8}$:
the limiting top performance is about 0.4~Eflop/s. Even for the system with the best parameters achieving 
that dream limit seems to be not realistic.

\begin{figure}
\begin{tikzpicture}[scale=.95]
\begin{axis}
[
	title={$R_{Max}$ of Top10 Supercomputers for benchmark \large $HPL$},
	width=\textwidth,
	cycle list name={my color list},
		legend style={
			cells={anchor=east},
			legend pos={north west},
		},
		xmin=0.005, xmax=1.1,% x scale
		ymin=0.003, ymax=0.3, % y scale
		xlabel={$R_{Peak}$  (exaFLOPS)},
		/pgf/number format/1000 sep={},
		ylabel={$R_{Max} (exaFLOPS)$},
		xmode=log,
		log basis x=10,
		ymode=log,
		log basis y=10,
		]
\addplot table [x=a, y=b, col sep=comma] {RMaxHPL10.csv};
		\addlegendentry{Taihulight}
\addplot table [x=a, y=c, col sep=comma] {RMaxHPL10.csv};
		\addlegendentry{Tianhe-2}
\addplot table [x=a, y=d, col sep=comma] {RMaxHPL10.csv};
		\addlegendentry{Piz Daint}
\addplot table [x=a, y=e, col sep=comma] {RMaxHPL10.csv};
		\addlegendentry{Gyoukou}
\addplot table [x=a, y=f, col sep=comma] {RMaxHPL10.csv};
		\addlegendentry{Titan}
\addplot table [x=a, y=g, col sep=comma] {RMaxHPL10.csv};
		\addlegendentry{Sequoia}
\addplot table [x=a, y=h, col sep=comma] {RMaxHPL10.csv};
		\addlegendentry{Trinity}
\addplot table [x=a, y=i, col sep=comma] {RMaxHPL10.csv};
		\addlegendentry{Cori}
\addplot table [x=a, y=j, col sep=comma] {RMaxHPL10.csv};
		\addlegendentry{Oakforest}
\addplot table [x=a, y=k, col sep=comma] {RMaxHPL10.csv};
		\addlegendentry{K computer}
%\addplot table [x=a, y=l, col sep=comma] {dat/RMaxHPL10.csv};
%		\addlegendentry{Mira}
		\addplot[only marks,  mark=o,  mark size=3,  thick] plot coordinates {
			(0.1254,0.09301) %Taihulight
			(0.0549,0.033863) %Tianhe-2
			(0.0253,0.01960) %Piz Daint
			(0.0282,0.01914) %Gyoukou
			(0.0271,0.01759) %Titan
		    (0.0201,0.01711) %Sequoia
		    (0.0439,0.01414) %Trinity
		    (0.0279,0.01401) %Cori
		    (0.0249,0.01355) %Oakforest
		    (0.0113,0.01051) %K computer
%		    (0.0100,0.00853) %Mira
		};
\end{axis}
\end{tikzpicture}
\caption{$R_{Max}$ performance of selected TOP10 (as of 2017 July) supercomputers
in function of their peak performance $R_{Peak}$, for the \gls{HPL} benchmark. The actual $R_{Peak}$ values are denoted by a bubble.}
\label{fig:exaRMax}
\end{figure}
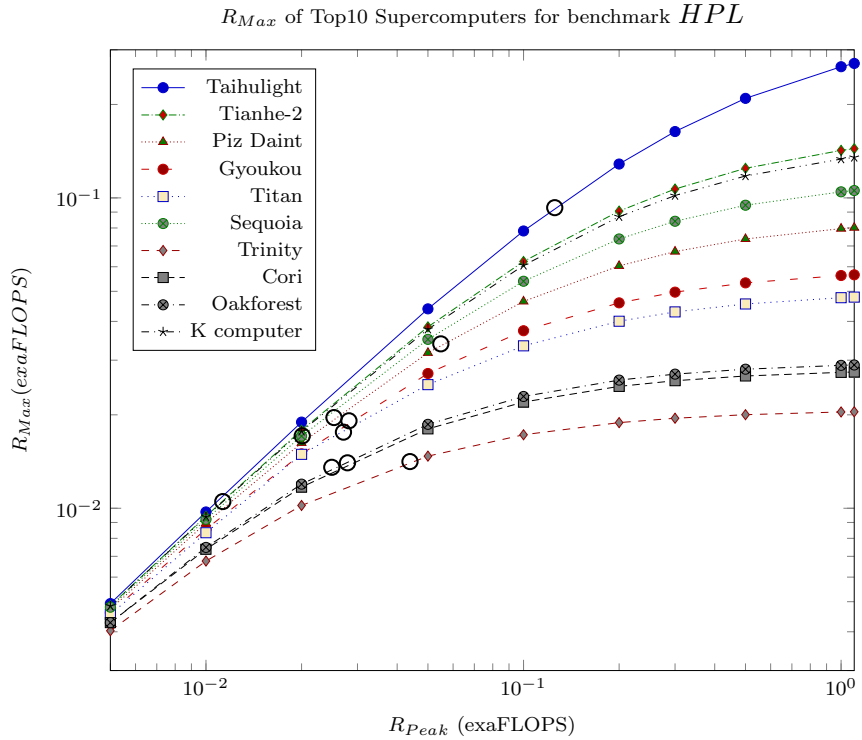

\begin{figure}
\begin{tikzpicture}[scale=.95]
\begin{axis}
[
	width=\textwidth,
	cycle list name={my color list},
		legend style={
			cells={anchor=east},
			legend pos={north west},
		},
		xmin=1e-6, xmax=0.5,% x scale
		ymin=1e-6, ymax=0.5, % y scale
		xlabel={$R_{Peak}$  (exaFLOPS)},
		/pgf/number format/1000 sep={},
		ylabel={$R_{Max} (exaFLOPS)$},
		xmode=log,
		log basis x=10,
		ymode=log,
		log basis y=10,
		]
\addplot table [x=a, y=h, col sep=comma] {RMaxvsRPeakatAlpha.csv};
		\addlegendentry{$1*10^{-8}$}
\addplot table [x=a, y=h, col sep=comma] {RMaxvsRPeakatAlpha.csv};
		\addlegendentry{$HPL$}
\addplot table [x=a, y=g, col sep=comma] {RMaxvsRPeakatAlpha.csv};
		\addlegendentry{$1*10^{-7}$}
\addplot table [x=a, y=f, col sep=comma] {RMaxvsRPeakatAlpha.csv};
		\addlegendentry{$1*10^{-6}$}
\addplot table [x=a, y=e, col sep=comma] {RMaxvsRPeakatAlpha.csv};
		\addlegendentry{$1*10^{-5}$}
\addplot table [x=a, y=d, col sep=comma] {RMaxvsRPeakatAlpha.csv};
		\addlegendentry{$HPCG$}
\addplot table [x=a, y=c, col sep=comma] {RMaxvsRPeakatAlpha.csv};
		\addlegendentry{$1*10^{-4}$}
\addplot table [x=a, y=b, col sep=comma] {RMaxvsRPeakatAlpha.csv};
		\addlegendentry{$3*10^{-4}$}
		\addplot[only marks,  mark=o,  mark size=5, very thick] plot coordinates {
			(0.125,0.093) %Taihulight at HPL
			(0.125,0.000375) %Taihulight at HPCG
		    (0.0113,0.0105) %K computer at HPL
		    (0.0113,0.0006) %K computer at HPCG
		};
\end{axis}
\end{tikzpicture}
\caption{$R_{Max}$ performance in function of peak performance $R_{Peak}$, at different $1-\alpha_{eff}()$ values. Bubbles 
display measured values
when using \gls{HPL} and \gls{HPCG} benchmarks, for $Taihulight$ and $K\ computer$, respectively.}
\label{fig:exaRMaxAlpha}
\end{figure}
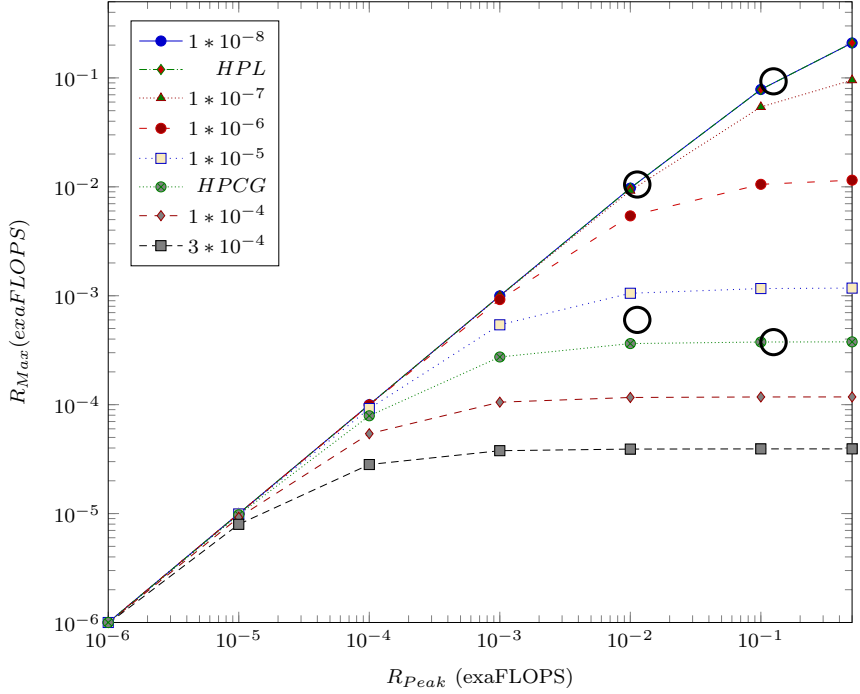

\subsection{Extrapolating the empirical parameters}
One way to derive more accurate estimations for the performance
limitations is to utilize the empirical model.
Keeping all other parameters constant, the number of processors can be virtually changed. 
Fig.~\ref{fig:exaRMax} depicts how the virtual versions of present TOP10 supercomputers will achieve the \textit{nominal} 1~Eflops/s.

To provide a feeling, how the effective parallelization influences the measurable performance,  Fig.~\ref{fig:exaRMaxAlpha} depicts what payload performance 
could be measured on that virtual Taihulight when running benchmark programs having different $(1-\alpha_{eff})$. 
This could be crucial when running real-life programs, the need for communication 
between processing units arises, and especially when they
must share some resource.

\subsection{Introducing a technical model}

Based on the empirical model, some technical meaning can be attributed to the $\alpha_{eff}^X$ components.
Although without considering the technical specifications in details, only the order of magnitude of the contributions can be estimated, it is accurate enough to draw some qualitative conclusions, especially of the limiting values of the different contributions.  The total $(1-\alpha_{eff})$ is about $3.3\times 10^{-8}$, so one upper limiting value  is known in advance: $(1-\alpha_{eff}^{SW})$ cannot be higher than that value. 

To turn our empirical model to a technical one, data published
in~\cite{DongarraSunwaySystem:2016} are used.
The 13,298 seconds benchmark runtime on the 1.45 GHz processors means 
$2*10^{13}$ clock periods. 
The absolutely necessary non-parallelizable activity is to start and stop the calculation. 
 If starting and stopping a program on a zero-sized supercomputer without \gls{OS} could be done in 2 clock periods,
then the absolute limit for $(1-\alpha)$ would be
$10^{-13}$.

From the model follows that two of the contributions 
can be critical when building "big" supercomputers.
The \gls{OS} looping contribution increases linearly with number of processors, and \gls{PD} contribution linearly increases with the
physical size of the computer.
As depicted in Fig.~\ref{fig:Ourmodel}, these contributions can be combined in such a way that small contributions from 
\gls{OS} are linked to large contributions from \gls{PD} and vice versa.
Anyhow, these two contributions will also provide an upper bound to
the absolute performance of supercomputers.
Since any of them can be quite small, the limit will be the
lower of the two individual bounds.

For considering \gls{PD} bound,
let us consider a cca. 100~meter sized computer having 1~GHz cores: the signal round trip time is cca. 
$10^{-6}$ seconds, or $10^{3}$ clock periods. When using high speed %100 Gbits/s
internal network, the message length has no considerable contribution
and a network message exchange time (including operating time of \gls{HW})
can be estimated to be of length $10^{-5}$ seconds, or $10^{4}$
clock periods.
So, the absolute limit for $(1-\alpha)$ of a supercomputer with
realistic size, but no operating system, is $10^{-9}$.

An operating system must, however, be used. If one considers context change with its consumed $10^4$ cycles~\cite{Tsafrir:2007},
the absolute limit is cca. $10^{-9}$, on a zero-sized supercomputer.
In addition, all cores must be manipulated through system calls, which contribution increases linearly with the number of cores and contribution from \gls{OS} can be dominant at high number of cores.

For the 10 million processors of Taihulight, at least $10^7$ clock cycles must be used.
Even when parameters can be passed in one clock cycle,   
for 10M parameter passings the absolute bound due to \gls{OS} looping contribution would be in the range of $10^{-6}$.
It is surely the dominating contribution for such large number of processors. 
Is then something wrong with the model? The measurable $(1-\alpha_{eff})$ for Taihulight must not be lower than any of the contributions, including the one due to looping in \gls{OS}.

At this point  one can understand the role of modularization some supercomputers utilize.
In the case of Taihulight, from the 260 cores 4 serves as “management processing element” (\gls{MPE})~\cite{FuSunwaySystem2016,DongarraSunwaySystem:2016},
so only the processors (or core groups) rather than individual cores shall be addressed, the rest will be organized by \gls{MPE}s.
This trick reduces the absolute computing performance of a processor only by 2\% on one side,
but on the other side reduces loop count by about two orders of magnitude, decreasing contribution $(1-\alpha_{eff}^{OS})$ by two orders of magnitude; in this way enabling to achive effective parallelization of value $1\times 10^{-8}$.
Just notice that the processors~\cite{CooperativeComputing2015} in $Taihulight$ attempt to reduce the non-payload time though utilizing special  \gls{OS} operating modes on the system, which enables application program to run without needing context change.

\subsection{Changing the computing model}

Introducing \gls{MPE}s decreased $(1-\alpha_{eff})$ and enabled
to build supercomputer with 10M processors and at the same time
reasonable efficiency.
Using \gls{MPE}s, however, violates computing paradigms: 
those "more equal" processors know that some other processors exist. 
As the above analysis demonstrated, (among others) the presently used 
\textit{Single-Processor Approach} (\gls{SPA}), that is \textit{the computing paradigm} itself, is a limiting factor
 in building larger supercomputers. 
The Explicitly Many-Processor Approach (\gls{EMPA})~\cite{VeghEMPAthY86:2016,IntroducingEMPA2018,RenewingComputingVegh:2017} enables to use forking-like handling of starting processing units,
and in this way the \gls{OS} looping contribution 
can be reduced from 10M cycles to 24,
in this way eliminating the most limiting obstacle from the way of building supercomputers from even more processors.

This is not against Amdahl's law: if the processors can cooperate, in Equ.~(\ref{eq:AmdahlBase}) $f(k)$ should be used instead of $k$, and the nature of $f(k)$ enables such drastic changes in the behavior of parallelly working systems.
It looks like Amdahl was right with saying:
\textit{"the organization of a single computer has reached its limits and that truly significant advances can be made only by interconnection of a multiplicity of computers in such a manner as to permit cooperative solution"}.

After introducing \gls{EMPA}, the context change becomes the largest contribution to $\alpha_{eff}^{OS}$.
Through introducing a reasonable layering~\cite{VeghLayering:2017}, this contribution can be lowered by orders of magnitude; making the propagation time \gls{PD} the dominating contribution.
It can be reduced by decreasing the physical size of supercomputers, say using 3D arrangement. Making all mentioned changes, in principle even Zflop/s supercomputers can be built. 
However, whithout making all those changes, even Eflop/s cannot be achieved. 

\section{Conclusions}\label{sec:summary}

The present technical implementations of supercomputers 
practically reached their technical limits.
The reliable database of parameters of supercomputers 
can be used to draw reliable statistical conclusions
on some parameters and limitations of supercomputers. 
Although the extrapolation of the tendencies enables
to make predictions for some future configurations,
the careful analysis reveals that the presently exclusively used Single-Processor Approach really forms an upper bound
for the performance of supercomputers. 
The experienced difficulties in building ever-larger supercomputers are of principial rather than technical nature.
%\bibliography{Bibliography.bib}   % name your BibTeX data base


\end{document}